\documentclass[%
reprint, superscriptaddress,
amsmath,amssymb, aps, pra,
floatfix, 
]{revtex4-1}
\usepackage{nicefrac}
\usepackage{graphicx}% Include figure files
\usepackage{dcolumn}% Align table columns on decimal point
\usepackage{bm}% bold math
\usepackage[protrusion=true,expansion=true,tracking=true]{microtype}
\usepackage{xcolor}
\usepackage{cancel}
\usepackage[utf8]{inputenc}

\usepackage{hyperref}% add hypertext capabilities
\usepackage[capitalize]{cleveref}

\crefname{section}{Sec.}{Sec.}

\DeclareMathOperator{\eul}{e}
\newcommand{\imag}{\ensuremath{\text{\textsl{i}}}}

\newcommand{\VC}{\hat{\vec{c}}}       % Fermion
\newcommand{\VCh}{\VC^\dagger}
\newcommand{\C}{\hat c}       % Fermion
\newcommand{\hamil}{\hat H}
\newcommand{\Ch}{\C^\dagger}
\newcommand{\U}{\hamil_\mathrm{int}}
\newcommand{\Uo}{\hamil_{\mathrm{int},o}}
\newcommand{\Ud}{\hamil_{\mathrm{int},d}}
\newcommand{\V}{\hat{V}}
\renewcommand{\vec}[1]{{\bm{#1}}}
\newcommand{\N}{\hat n}

\newcommand{\hc}[0]{\ensuremath{\mathrm{H.c.}}}
\newcommand{\mycaption}[2]{\caption[#1]{\emph{#1} #2}}

\begin{document}
\title{Topological phases in the Fermi-Hofstadter-Hubbard model on hybrid-space ladders}
\author{L.~Stenzel}
\affiliation{Department of Physics,
	Arnold Sommerfeld Center for Theoretical Physics (ASC),
	Ludwig-Maximilians-Universit\"{a}t M\"{u}nchen,
	D-80333 M\"{u}nchen, Germany.}
\affiliation{Munich Center for Quantum Science and Technology (MCQST), Schellingstr. 4, D-80799 M\"unchen, Germany}
\author{A.~L.~C.~Hayward}
\affiliation{Institute for Theoretical Physics, Georg-August-Universit\"at G\"ottingen,Friedrich-Hund-Platz 1, D-37077 G\"ottingen, Germany}
\author{U.~Schollw\"ock}
\affiliation{Department of Physics,
	Arnold Sommerfeld Center for Theoretical Physics (ASC),
	Ludwig-Maximilians-Universit\"{a}t M\"{u}nchen,
	D-80333 M\"{u}nchen, Germany.}
\affiliation{Munich Center for Quantum Science and Technology (MCQST), Schellingstr. 4, D-80799 M\"unchen, Germany}
\author{F.~Heidrich-Meisner}
\email{heidrich-meisner@uni-goettingen.de}
\affiliation{Institute for Theoretical Physics, Georg-August-Universit\"at G\"ottingen,Friedrich-Hund-Platz 1, D-37077 G\"ottingen, Germany}
\date{\today}

\begin{abstract}
In recent experiments with ultracold atoms,
both two-dimensional (2d) Chern insulators and one-dimensional (1d) topological charge pumps have been realized.
Without interactions,
both systems can be described by the same Hamiltonian, 
when some variables are being reinterpreted.
In this paper,
we study the relation of both models when Hubbard interactions are added,
using the density-matrix renormalization-group algorithm.
To this end,
we express the fermionic Hofstadter model in a hybrid-space representation,
and define a family of interactions,
which connects 1d Hubbard charge pumps to 2d Hubbard Chern insulators.
We study a three-band model at particle density $\rho=2/3$, 
where the topological quantization of the 1d charge pump changes from Chern number $C=2$ to $C=-1$ as the interaction strength increases.
We find that the $C=-1$ phase is robust when varying the interaction terms on narrow-width cylinders.
However, this phase does not extend to the limit of the 2d Hofstadter-Hubbard model,
which remains in the $C=2$ phase.
We discuss the existence of both topological phases for the largest cylinder circumferences
that we can access numerically.
We note the appearance of a ferromagnetic ground state between the strongly interacting 1d and 2d models.
For this  ferromagnetic state,
one can understand the $C=-1$ phase from a bandstructure argument.
Our method for measuring the Hall conductivity could similarly be realized in experiments:
We compute the current response to a weak, linear potential,
which is applied adiabatically.
The Hall conductivity converges to integer-quantized values for large system sizes,
corresponding to the system's Chern number.
\end{abstract}

\maketitle
\section{Introduction}
In the last years,
various systems with topological properties have been realized in experiments with ultracold atomic gases in optical lattices \cite{Goldman2014,Aidelsburger2018,Cooper2019}.
In lattices with two spatial dimensions,
both the Hofstadter model \cite{Miyake2013,Aidelsburger2013}
and the Haldane model \cite{Jotzu2014,Tarnowski2019} have been realized.
The Hofstadter model has also been implemented in synthetic dimensions \cite{Stuhl2015,Mancini2015,An2017,Kolkowitz2017},
where the spin degree of freedom is interpreted as sites on the rungs of a ladder model.

As pointed out by \citet{Thouless1983},
quantum Hall physics can also be observed in a family of one dimensional models.
Diagonalizing a two-dimensional (2d) quantum Hall model along one spatial dimension, 
the resulting quasimomentum can be interpreted as the pump parameter.
The number of charges transported in each adiabatic pump cycle is then quantized by the Chern number.
Such charge pumps have also been realized in ultracold-atom experiments \cite{Lohse2016,Nakajima2016,Lohse2018}.

Studying the effects of particle interactions in all of these experiments remains challenging:
While on-site interactions are typically present with ultracold atoms \cite{Bloch2008},
the experiments have been performed in the limit of either vanishing or hard-core interactions.
Accessing strong but finite interactions 
and reaching the low-filling regime remains elusive
due to heating \cite{Jotzu2014,Reitter2017},
except for the few-body limit \cite{Tai2017} or certain 1d systems \cite{Lohse2016,deLeseleuc2019}.

In a previous work \cite{Stenzel2019},
we showed that Hubbard interactions in a fermionic, one-dimensional charge pump can change its topological properties:
Without interactions, the Chern number is $C=2$,
but strong repulsion changes it to $C=-1$.
Note that we change the sign convention for $C$ relative to \cite{Stenzel2019}.
This topological transition is related to a series of two 1d quantum phase transitions,
which occur for certain values of the pump parameter.
For these configurations,
the 1d charge pump corresponds to the three-site ionic Hubbard model \cite{Torio2006,Fabrizio1999,Yamamoto2001,Egami1993,Manmana2004,Torio2001,Otsuka2005,Kampf2003,Batista2004,Lou2003,Aligia2004,MurciaCorrea2016}.
The interaction-driven change of topological properties in charge pumps with either fermions or bosons has also been studied in related, earlier papers \cite{Nakagawa2018,Hayward2018,Hu2019}. 

The analytic one-to-one correspondence of charge pumps and 2d quantum Hall models breaks down when interactions are introduced.
In this paper,
we study numerically
whether 1d charge pumps 
and 2d quantum Hall models with Hubbard interactions are adiabatically connected.
In particular,
we try to find a phase with Chern number $C=-1$ in the Hubbard-Hofstadter model,
which is adiabatically connected to the $C=-1$ interacting charge pump described in \cite{Stenzel2019}.

Our starting point is the Harper-Hofstadter model \cite{Harper1955,Hofstadter1976},
which is a paradigmatic model for studying the quantized Hall conductivity in a lattice \cite{Thouless1982}.
We study the three-band Hofstadter model with two spinful fermions per every three lattice sites,
corresponding to parameters chosen previously \cite{Stenzel2019}.
In order to connect 1d and 2d physics,
we express the model in a mixed real- and momentum-space representation, 
called hybrid space \cite{Ehlers2017,Motruk2016}.
In hybrid space, we can tune the interactions in such a way
that the 1d Hubbard charge pump and the 2d Hubbard-Hofstadter model become the limiting cases.

Numerically,
we are restricted in the lattice sizes we can study.
Increasing the system's width is much more expensive than its length.
In hybrid space, 
we can use a cylindrical geometry without increasing the numerical cost \cite{Motruk2016,Ehlers2017}.
We use twisted boundary conditions along the width 
and average over multiple twist angles to reduce finite-size effects.

We compute the Hall conductivity
by measuring persistent currents as a response to an adiabatically applied linear potential.
We observe a finite Hall conductivity in insulating phases,
which converges to integer values as we increase system size.
We identify topological phases with two non-zero Chern numbers.

Several experiments with ultracold atoms and artificial gauge fields have already measured the response to an external, linear potential \cite{Aidelsburger2015,Stuhl2015,Mancini2015,Genkina2019}.
There are different theoretical proposals to measure Chern numbers in such setups using bosonic wave packets \cite{Price2012,GoldmanX2014,Mugel2017}
or fermionic systems \cite{Dauphin2013} under the action of a constant force.
A  method for measuring non-quantized Hall responses in interacting lattice models that is similar to ours has recently been proposed \cite{Greschner2019}.

We find that the $C=-1$ phase exists in large regions in our space of interaction parameters.
However, our results suggest that the Hofstadter-Hubbard model remains adiabatically connected to the band-insulating phase with a Chern number $C=2$, even for large interactions.
Thus, the 1d and 2d limits would be separated by a topological transition.
Most results are obtained in the narrow-cylinder limit of width $W=2$. 
We discuss the existence of the strongly interacting $C=-1$ phase for wider systems, up to $W=6$.

Finally, we discuss the appearance of a ferromagnetic (FM) ground state for some interaction parameters inside the $C=-1$ phase.
The FM phase exists for all system sizes we consider,
but does not extend to the 2d or 1d limit.
A FM state has the Chern number $C=-1$ since the system is then equivalent to free spinless fermions.

The paper is structured in the following way:
In \cref{sec:model}, we describe our model and 
explain how it relates to both the 2d Hubbard-Hofstadter model and to
interacting 1d superlattice charge pumps.
The following \cref{sec:method_obs} briefly describes our numerical methods 
and discusses the observables used in this paper. 
Section \ref{sec:jy_s0} discusses the Hall conductivity depending on interaction parameters of the model.
We reproduce the topological transition of the 1d charge pump in \cref{sec:1dlimit}
and study the extended parameter space in the numerically accessible regime of a small system width in \cref{sec:w2}.
In \cref{sec:width_scaling}, we show  that both topological phases persist for wider systems.
In \cref{sec:fm_gs}, we discuss the ferromagnetic ground state,
which exists for some interaction parameters.
We conclude with a summary in \cref{sec:summary}.
%\hl{
	\Cref{sec:app_chi} contains data for the Hall response at additional interaction strengths.
	In \cref{sec:e_v_fit}, we discuss the numerical accuracy of our data.
	We show the time-dependent Hall response induced by a quenched external potential in \cref{sec:quench_dynamics}.
	\cref{sec:S_sectors} contributes to the discussion of \cref{sec:fm_gs} and contains additional data for the dependence of ground-state properties on the total spin.
%}

\section{Fermi-Hofstadter-Hubbard model}
\label{sec:model}

\begin{figure}[tb] 
    \centering \includegraphics[width=\columnwidth, clip]{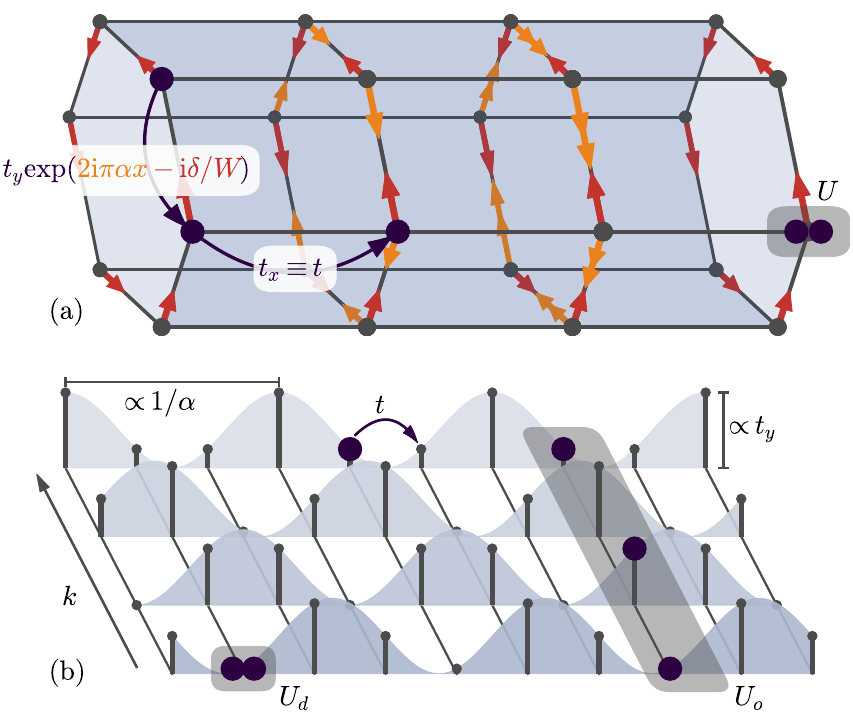}
    \mycaption{Sketch of the Hofstadter-Hubbard model.}{
        \textbf{(a)} \textbf{Real-space} representation on a cylindrical geometry,
        with a twist angle $\delta$ implemented via homogeneous, complex hopping rates along the $y$ direction. 
        The magnetic field is implemented via position-dependent complex phases, 
        sketched here for flux $\alpha=1/3$.
        The interaction is proportional to $U$ and is purely on-site.
        \textbf{(b)} \textbf{Hybrid-space} representation,
        obtained via a Fourier transformation along the axis with periodic boundary conditions (see also Ref. \cite{Saito2017}).
        The flux $\alpha=1/3$ corresponds to a three-site periodic superlattice potential and
        its amplitude is shown in light blue.
        The interaction is now delocalized over each ring: 
        We split the terms $\U = \Uo + \Ud$ according to \cref{eq:split_int} into terms which are diagonal (on-site) in the hybrid-space basis, $\Ud$, and the rest,
        which is off-diagonal (ring-wise) in hybrid space, $\Uo$.
        The interaction terms have strengths $U_d$ and $U_o$, respectively.
        The model sketched in (b) maps to (a) only when $U=U_d=U_o$.
        Note that the number of sites along each dimension is of course preserved when going from (a) $\to$ (b). 
        A different number of sites was chosen in (a) and (b) for visualization purposes.
    }
    \label{fig:model_sketch}
\end{figure}

The Hofstadter-Hubbard Hamiltonian for spinful fermions, $\sigma\in\{\downarrow,\uparrow\}$,
on a cylinder of length $L$ and circumference $W$ can be written as,
\begin{equation}
\label{eq:fhhrs}
\begin{aligned}
\hamil=\sum_{x=1}^L\sum_{y=1}^W\bigg[&
\begin{aligned}[t]
\sum_{\sigma}\bigg(&-t_y\eul^{2\pi\imag \alpha x-\imag\delta/W}\Ch_{x,y,\sigma}\C_{x, y+1,\sigma} \\
&-t\Ch_{x,y,\sigma}\C_{x+1,y,\sigma} +\hc\bigg)        
\end{aligned}\\ &+U\N_{x,y,\uparrow}\N_{x,y,\downarrow}\bigg]\,.
\end{aligned}
\end{equation}
The boundary conditions are implemented via
$\C_{L+1,y,\sigma}\equiv 0$ and $\C_{x,W+1,\sigma}\equiv \C_{x,1,\sigma}$.
The on-site Hubbard repulsion is of strength $U$.
The model is sketched in \cref{fig:model_sketch}(a).
The hopping term along the ring includes a complex phase:
A particle hopping around one plaquette gains a phase $\alpha$,
corresponding to a magnetic flux piercing each plaquette.
In this paper, we only consider the case of $\alpha=1/3$,
i.e., one flux quantum per three lattice sites.
We choose this value of the flux
because $3$ is the smallest integer denominator for which the Hofstadter model exhibits topologically nontrivial bands \cite{Thouless1982}.
There is also a flux $\delta$ piercing the cylinder along its height, 
which we interpret as an angle twisting the boundaries.
Twist angles can be used to define many-body topological invariants \cite{Niu1985}.
We will average over $\delta$ to reduce the effects of a finite width $W$.

For the rest of this paper, we study the phases at fixed particle density $\rho=2/3$, 
i.e., two spin-$1/2$ fermions per every three sites.
For $\alpha=1/3$ and in the free case $U=0$, 
this corresponds to a band insulator with Chern number $C=2$,
as the lowest band has $C=1$ and is filled by both spin species.
We choose anisotropic tunneling rates $t_y=1.5t$ such that the parameters correspond to the charge pump considered before \cite{Stenzel2019}.

\subsection{Hybrid-space representation}
\label{sec:hybrid_space}
By Fourier transforming \cref{eq:fhhrs} along the periodic $y$-axis, 
we find a mixed real- and momentum-space representation, 
which we call hybrid space,
\begin{equation}
\label{eq:hshh}
\hamil=\sum_{x,k,\sigma}
\begin{aligned}[t]
\bigg[&-2t_y\cos(2\pi(\alpha x+k/W)-\delta/W)\N_{x,k,\sigma}\\
&-t \Ch_{x,k,\sigma}\C_{x+1,k,\sigma}
\bigg] + \U\,.
\end{aligned}
\end{equation}
The hybrid-space model is sketched in \cref{fig:model_sketch}(b).
Not taking $\hamil_{\rm int}$ into account,
\cref{eq:hshh} can be understood as a set of uncoupled 1d chains,
which are labeled by quasimomentum $k$.
There is an additional cosinusoidal potential depending on $k$, a superlattice.

In the case of a strictly 1d charge pump, $W=1$,
the topology of the Hofstadter bandstructure manifests itself by an integer-quantized amount of charges transported in each adiabatic pump cycle $\delta\to\delta+2\pi$ \cite{Thouless1983}.

In hybrid space, 
the onsite Hubbard repulsion becomes delocalized over each ring,
\begin{equation}
\label{eq:split_int}
\begin{aligned}
\U =&\, \frac{U}{2}\sum_{x,y} \VCh_{x,y}\cdot \VC_{x,y}\left(\VCh_{x,y}\cdot \VC_{x,y}-1\right)\\
=&\,\frac{U}{2W}\sum_x\sum_{k,p,q}\VCh_{x,k}\cdot\VC_{x,p} \times \VCh_{x,q}\cdot\VC_{x,k+q-p} \\ &\,- \frac{U}{2}\sum_{x,k}\N_{x,k}\\
=:&\, U_{d}\Ud + U_{o}\Uo\,,
\end{aligned}
\end{equation}
where we use spinor operators, $\VC=(\C_\uparrow,\C_\downarrow)^T$ to simplify the notation.
In the last line, we split the interaction into two parts:
$\Ud$ contains contributions that are diagonal in the hybrid-space indices $x,k$.
All remaining, off-diagonal terms are grouped in $\Uo$, 
which is delocalized over each ring.
Note that terms proportional to the total particle number $\sum_{x, k} \N_{x,k}$ only shift the chemical potential
and can be neglected when the particle number is fixed by the numerical method.
Explicitly, the interaction terms take the following form,
\begin{align}
\label{eq:ud_uo}
\Ud :=&\, \frac{1}{2W}\sum_{x,k} \N_{x,k}(\N_{x,k}-1)\,,\\
\Uo :=&\, 
\frac{1}{2W}\sum_{x,k}\bigg(
\begin{aligned}[t] &\sum_{p,q}(1-\delta_{k,p}\delta_{k,q})\VCh_{x,k}\cdot\VC_{x,p}\\
&\times\VCh_{x,q}\cdot\VC_{x,k+q-p} -(W-1)\N_{x,k}\bigg)\,.
\end{aligned}
\end{align}
The term $\Ud$ looks like the normal Hubbard interaction, 
scaled by $W^{-1}$.
This term thus corresponds to the 1d interaction in a charge pump as
$U_d=W\,U_{1d}$.

\begin{figure}[tb] 
	\centering \includegraphics[width=\columnwidth, clip]{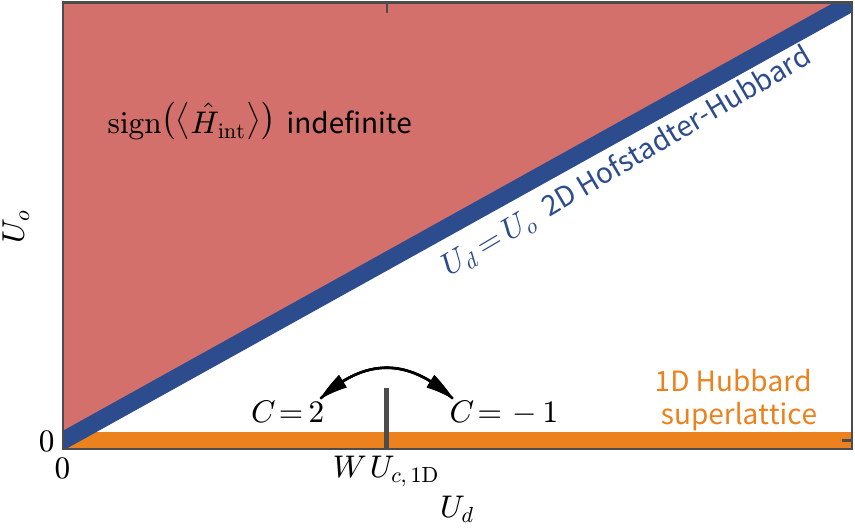}
	\mycaption{Sketch of the parameter space created by splitting the interaction term according to \cref{eq:split_int}.}{
	At $U_o=0$, the interaction is onsite in the hybrid-space representation, 
	such that the system consists of uncoupled Hubbard chains with a periodic potential.
	For the 1d model, there is a critical interaction strength $U_{\rm c,1d}$,
	where the Chern number changes from $C=2$ to $C=-1$ \cite{Stenzel2019}.
	Due to the prefactor in \cref{eq:ud_uo}, 
	we obtain a factor of $W$ for the critical value of $U_d$.
	When we fix $U=U_d=U_o$, we recover the original 2d Hubbard interaction.
	Note that both $\Ud$ and $\Ud+\Uo$ are positive semidefinite,
	however $\Uo$ is not.
	Therefore, the interaction can be attractive for $U_o>U_d$, which we do not consider in this paper.
	}
	\label{fig:ud_uo_parameter_space}
\end{figure}

The parameterization of \cref{eq:ud_uo} allows us to relate 1d charge pumps with interactions ($U_o=0,\,U_d>0$) to the interacting 2d Hofstadter model ($U_d=U_o>0$),
as sketched in \cref{fig:ud_uo_parameter_space}.
In this figure,
these limiting cases are represented by the blue and orange lines.
Note that while both $\U=\Ud+\Uo$ and $\Ud$ are positive semidefinite,
$\Uo$ is not.
Thus, for $U_o>U_d$, the interactions can become attractive 
and we do not consider this case in this paper.

As shown in \cref{fig:ud_uo_parameter_space}, 
there is a topological phase transition from Chern number $C=2$ to $C=-1$ for $U_o=0$ and a critical interaction strength $U_d=W\,U_{c,{\rm 1d}}$,
corresponding to uncoupled 1d superlattice chains.
We studied this 1d phase transition in the context of charge pumps in a previous paper \cite{Stenzel2019}.
We expect weakly-interacting systems with parameters $U_d,U_o\ll W\cdot U_{c,{\rm 1d}}$ 
to be adiabatically connected to the free model,
and thus to have Chern number $C=2$.

	For the strongly-interacting 1d charge pump with Chern number $C=-1$,
	both bulk and spin gaps vanish for certain values of the pump parameter \cite{Stenzel2019}.
	This corresponds directly to the gap closing in the  ionic Hubbard model \cite{Fabrizio1999,Manmana2004}.
	While the system remains insulating, i.e., the charge gap remains open,
	the	topological quantization could, in principle, break down as perturbations are added.
	Here, we want to find out whether the $C=-1$ phase obtained in the 1d limit, $U_d>W\,U_{c,{\rm 1d}}$,
	also exists with 2d interactions, $0<U_o\leq U_d$.

\section{Methods and observables}
\label{sec:method_obs}

\subsection{Methods}
\label{sec:method}
All numerical results presented in this paper are obtained using the density-matrix renormalization-group (DMRG) algorithm \cite{White1992,Schollwoeck2011}. 
We employ a single-site variant \cite{Hubig2015} of this algorithm, 
as implemented in the SyTen toolkit \cite{SyTen,Hubig2017}.
DMRG is a method for 1d systems, however,
one can map \cref{eq:hshh} onto a $W\times L$ sites-long 1d chain.
Any lattice site, labeled by $x$ and $k$, 
is mapped onto a position $i$ via $i = W\cdot x+k$ in a matrix-product state (MPS).
This mapping introduces long-range correlations in the 1d description,
generally increasing the computational cost exponentially in $W$ \cite{Stoudenmire2012}.

In DMRG, we enforce the conservation of particle number $U(1)$ and spin $SU(2)$ symmetry.
Furthermore, we use the $k$ labels introduced in \cref{sec:hybrid_space} to fix the $Z_w$ symmetry sector of total quasimomentum along the $y$-axis.
For all parameters considered, the lowest energy state is in the $K:=\sum_{x,k}k\langle\N_{x,k}\rangle =0\, (\mathrm{mod}\; W)$ sector.
We fix particle density to $\rho:=N/(W L)=2/3$ and total spin to be $S=0$.

Large bond-dimensions $m$ of the MPS are required for convergence, 
especially with off-diagonal interactions, $U_o\gg t$:
We use $m_{SU(2)}=8,\dots,12 \times 10^3$ which would
correspond to $m_{U(1)}=2,\dots, 10 \times 10^4$, 
when only enforcing the Abelian spin $S^z$ symmetry.
%\hl{
The ratio $m_{U(1)}/m_{SU(2)}$ 
at a given MPS bond depends on the occupation of higher spin multiplets, 
due to their $2S+1$ fold degeneracy.
It varies with model parameters
and $m_{U(1)}/m_{SU(2)} \gtrsim 10$ is particularly large in the region discussed in \mbox{\cref{sec:fm_gs}}.
%}

Computing the error of a DMRG result can be more expensive than the ground-state search itself.
We use the two-site variance of the Hamiltonian $\mathrm{var}_2(\hamil)$ 
as a measure of DMRG convergence \cite{Hubig2018}.
Especially for 2d models, 
this approximation is much cheaper than computing the full variance.
However, $\mathrm{var}_2(\hamil)$ is still too expensive for the largest systems and bond-dimensions used here and in those case,
we rely on studying the observables as a function of  bond dimension. 

Studying short $L=12$ systems of width $W=3$ with bond dimensions up to $m_{SU(2)}=4000$,
we find a strong dependence of $\mathrm{var}_2(\hamil)$ on system parameters.
In many cases, $m_{SU(2)}= 500$ is sufficient to reach $\mathrm{var}_2(\hamil)<10^{-6}t^2$,
but there are also parameters for which $m_{SU(2)}= 4000$ only yields $\mathrm{var}_2(\hamil)<10^{-3}t^2$.
Extrapolating the energy in $\mathrm{var}_2(\hamil)$ \cite{Hubig2018},
it seems that for these models and parameters, 
the error in the  energy is on the same order as the two-site variance, $E_\mathrm{DMRG}-E_\mathrm{exact}=\mathcal{O}(\mathrm{var}_2(\hamil)/t)$.

Since we use much higher bond dimensions for longer systems,
we are confident in the accuracy of our results for narrow cylinders $W=2,3$.
For the largest cylinders of width $W=5,6$, 
errors are certainly larger and
in these systems, we might not capture the position of the topological phase transition accurately.
However, we can still find phases with different signs of $\chi_{\rm Hall}$, 
consistent with data for narrow systems.

To  access the quality of the numerical data,
we compare DMRG results for different initial states and different parameters.
In particular, we apply a weak linear potential $V$, 
as described below in \cref{sec:hall}, 
and verify the linear behavior of $E_{\rm gs}(V)$, 
see \cref{sec:e_v_fit}.

In \cref{sec:fm_gs}, we also compute the energy of the ferromagnetic state, $S=N/2$.
Due to the Pauli principle, double occupation is prohibited both in real and hybrid space.
Therefore, both $\langle\Ud\rangle$ and $\langle\Ud+\Uo\rangle$ vanish
and it is sufficient to solve the noninteracting Hamiltonian, 
which does not require DMRG.

\subsection{Hall current}
\label{sec:hall}
In the first part of this section, 
we describe our setup for computing the Hall response 
and define the observables.
Then, we show how these measurements can be related to topological quantization
for simulations performed in finite-size systems.

We use a method to compute the Hall conductivity, 
which could very similarly be realized in experiments with cold atoms.
In order to probe the Hall current, 
we add a weak ($V\ll t$) linear potential to the Hamiltonian. 
The potential is constant along the $y$ (equivalently: $k$) direction
and increases linearly along the $x$ direction,
\begin{equation}
\label{eq:Vgrad}
\V = V \sum_{x,k,\sigma} x\,\N_{x,k,\sigma}\,.
\end{equation}
This corresponds to a constant electric field along the $x$-direction.
We can apply $\V$ exactly adiabatically by performing consecutive ground-state DMRG runs for different field strengths $V$.

Eventually, we are only interested in the limit $V\to 0$ in order to stay in the regime of a linear Hall response \cite{Kubo1957}.
In our simulations, we consider $5$ to $10$ different values of $V$ in the range  $0\leq V\leq 0.1,\dots, 1 \times 10^{-2}t$ and fit a linear function to the computed currents.
A larger number of different potential strengths improves our estimate of the fit's accuracy.
The range of $V$ for which a linear behavior is observed depends on the size of the system and the many-body gaps.

\begin{figure}[tb] 
    \centering \includegraphics[width=\columnwidth, clip]{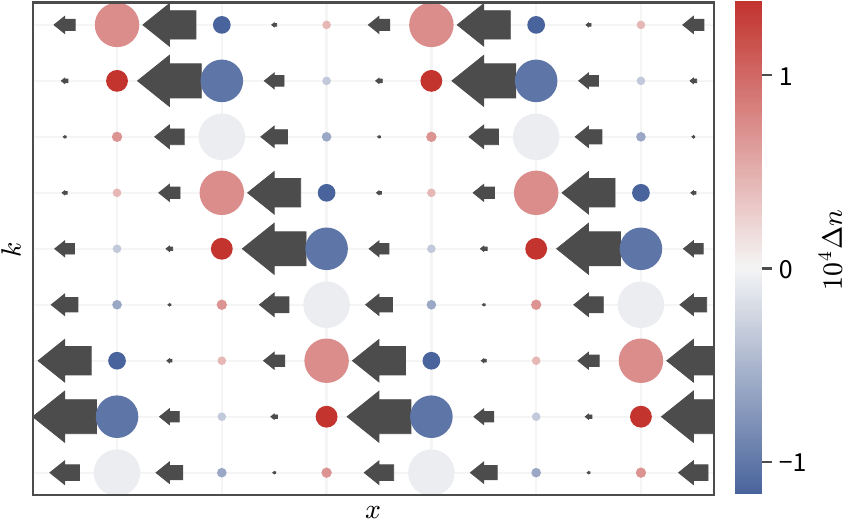}
    \mycaption{Response to a potential gradient $\V$ in hybrid space.}{
        We show the system at a short time $\tau\, t=0.2$ after switching on a potential gradient of strength $V=0.01t$: 
        We use a time-dependent simulation for this illustration as
        there are no currents along the $x$  direction in the ground state on a cylinder.
        Data are shown for the bulk of a noninteracting system of size $W=9,\,L=60$, with twist angle $\delta=0.2\pi$.
        The size of an arrow indicates the amplitude of the particle current in the $x$ direction,
        the size of the circles encodes the occupation number $\langle\N_{x,k}\rangle$ on a lattice site.
        Colors indicate the particle number difference compared to before the quench,
        $\Delta n=\langle \N (\tau)\rangle - \langle \N(0)\rangle$.
        Note that in the free model, 
        the hybrid-space legs are uncoupled and 
        the current $\langle \hat j_y(x,k)\rangle$ appears as a quasi-momentum $k$ dependent polarization along the $x$ direction. 
    }
    \label{fig:jy_in_hybrid_space}
\end{figure}

The cylindrical geometry sketched in \cref{fig:model_sketch}(a) allows for persistent ground-state currents along the rings.
Taking the twist angle $\delta$ and anisotropic tunneling rates into account,
we can express the intra-ring current as
\begin{equation}
\label{eq:hsjy}
\begin{aligned}
\hat{j}_y(x) :=&\, \frac{\imag t_y }{W}\sum_{y,\sigma} \eul^{2\pi\imag \alpha x-\imag\delta/W}\Ch_{x,y,\sigma}\C_{x,y+1,\sigma} + \hc\\
=&\,\frac{2 t_y}{W}\sum_{k,\sigma} \sin(2\pi(\alpha x+k/W)-\delta/W)\N_{x,k,\sigma}\,.
\end{aligned}
\end{equation}
Note that in the hybrid-space representation, 
$\hat{j}_y$ is a sum of operators acting on a single site.
This is related to the fact that the legs in the free hybrid-space Hamiltonian given in \cref{eq:hshh} are not coupled.
The Hall-current response to $V\neq 0$ is thus due to a polarization \emph{along} the direction of the potential gradient,
which depends on $k$ and $x$.
This is sketched in \cref{fig:jy_in_hybrid_space}:
In response to a weak potential $\V$, which is switched on instantaneously,
particles hop along the $x$ direction in such a way
that a Hall current $\langle \hat{j}_y \rangle$ as defined in \cref{eq:hsjy} is created.
We choose a quench for \cref{fig:jy_in_hybrid_space} because there are no currents along the $x$ direction in the ground state of an open system.

We define the linear Hall response to a weak potential gradient as
\begin{equation}
\label{eq:chi_hall}
\chi_{\rm Hall}:= 2\pi\, \partial_V \left.\left\langle \hat{j}_y(x)\right\rangle_{x\in \mathrm{bulk}}\right|_{V\to 0}\,,
\end{equation}
where we restrict the average to rings in the bulk of the cylinder.
In most cases, 
we find it sufficient to ignore $3$ or $6$ rings on either end of the cylinder,
in order to observe bulk behavior.

\subsubsection{Quantized Hall response}
\label{sec:quantized_chi_hall}
The Hall response defined in \cref{eq:chi_hall} can be computed in any interacting, 
finite-size system, but does not take integer values,
which one would like to see for topologically quantized systems.

To define the Chern number for a finite, interacting model,
one usually employs twisted boundaries for both spatial dimensions to define the Berry curvature on the parameter space of twist angles \cite{Niu1985}.
This approach is commonly used with numerical methods to compute exactly integer-quantized Chern numbers from a finite number of finite-size ground states \cite{Fukui2005}.
Previously, we have also used this method in the limit of 1d systems \cite{Stenzel2019}.

To recover the integer quantization of the Hall response $\chi_{\rm Hall}$,
we need to average over the twist angle $\delta$
\begin{equation}
\left\langle \chi_{\rm Hall}\right\rangle_{\delta} = C\in\mathbb{Z}\,.
\end{equation}

\begin{figure}[tb] 
	\centering \includegraphics[width=\columnwidth, clip]{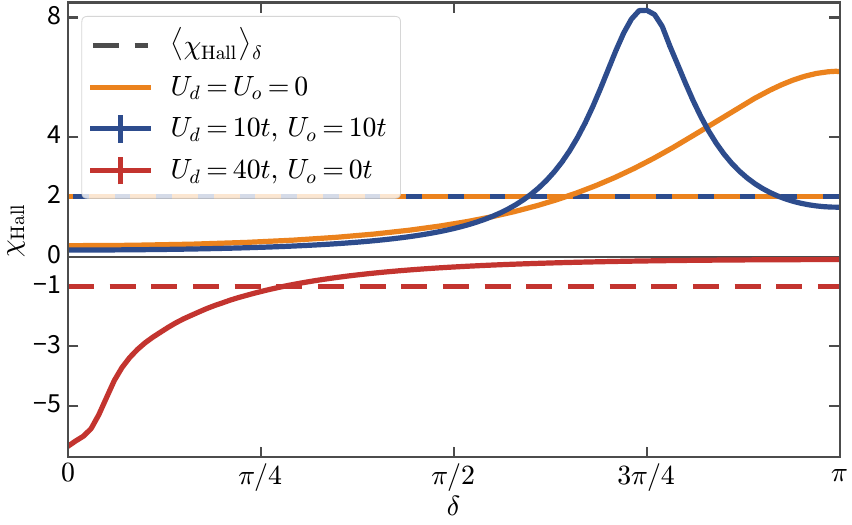}
	\mycaption{Dependence of the Hall susceptibility on the twist angle $\delta$.}{
    Data are shown for a narrow cylinder of $W=3$ and $L=24$.
    $\chi_{\rm Hall}$ is extracted from a linear fit as the potential $\V$ is applied adiabatically.
    The Hall response $\chi_{\rm Hall}$ 
    depends strongly on the twist angle $\delta$ and interaction parameters.
    Only the $\delta$ average is integer quantized:
    The dashed lines indicate the corresponding averages and assume integer values up to finite-size effects, which are on the order of $10^{-3}$.
    We observe that for each parameter combination of $U_d$ and $U_o$, 
    the response $\chi_{\rm Hall}$ has the same sign for all values of $\delta$.
	}
	\label{fig:jy_quantization}
\end{figure}

We show the dependence of the Hall response $\chi_{\rm Hall}$ on the twist angle $\delta$ for different interaction strengths in \cref{fig:jy_quantization}.
The amplitude of $\chi_{\rm Hall}$ depends strongly on $\delta$ for the narrow width $W=3$ considered here.
Computing the average over $\delta$,
we recover integer values for $\langle\chi_{\rm Hall}\rangle_\delta$, 
up to a precision of $5\cdot 10^{-3}$.
We found a discrepancy of the same order when studying finite, open chains of similar length $L$ \cite{Stenzel2019}.

Compared to the method by \citet{Fukui2005} to numerically integrate the Berry curvature,
our $\delta$ average does not give integer values \emph{by design}.
Instead, we may converge to a integer as the number of samples 
and the system size increases.
We expect this to happen if and only if the system is in a topologically nontrivial,
insulating phase.

\section{Hall conductivity in the \texorpdfstring{$S=0$}{S=0} ground state at \texorpdfstring{$\rho=2/3$}{ρ=2/3}}
\label{sec:jy_s0}
In this section, 
we study the adiabatic Hall response $\chi_{\rm Hall}$ to a weak gradient \cref{eq:Vgrad} for different parameters of the model \cref{eq:hshh}.
We restrict the DMRG ground-state search to the spin-singlet sector, $S=0$.

This section is structured as follows:
In \cref{sec:1dlimit}, we reproduce the 1d topological phase transition \cite{Stenzel2019}. 
Specifically, we run simulations for width $W>1$, 
but fix the off-diagonal interaction strength to $U_o=0$.
In \cref{sec:w2}, we extend the parameter space to $U_o\leq U_d$,
but  restrict ourselves to width $W=2$.
Finally, in \cref{sec:width_scaling}, 
we present data for wider cylinders and $U_o\neq 0$,
and discuss how critical interaction strengths scale with the width.

\subsection{Quasi 1d limit \texorpdfstring{$U_o=0$}{Uo=0}}
\label{sec:1dlimit}
For $U_o=0$, 
\cref{eq:hshh} can be interpreted as a series of $W$ uncoupled 1d superlattices 
with different superlattice phases $\delta$.
In this section, we verify that computing $\chi_{\rm Hall}$ reproduces the topological transition that we discussed in a previous paper \cite{Stenzel2019}.
Unlike for chains, we do not keep particle numbers on each leg fixed individually, 
which could in principle yield a different behavior.

\begin{figure}[tb] 
	\centering \includegraphics[width=\columnwidth, clip]{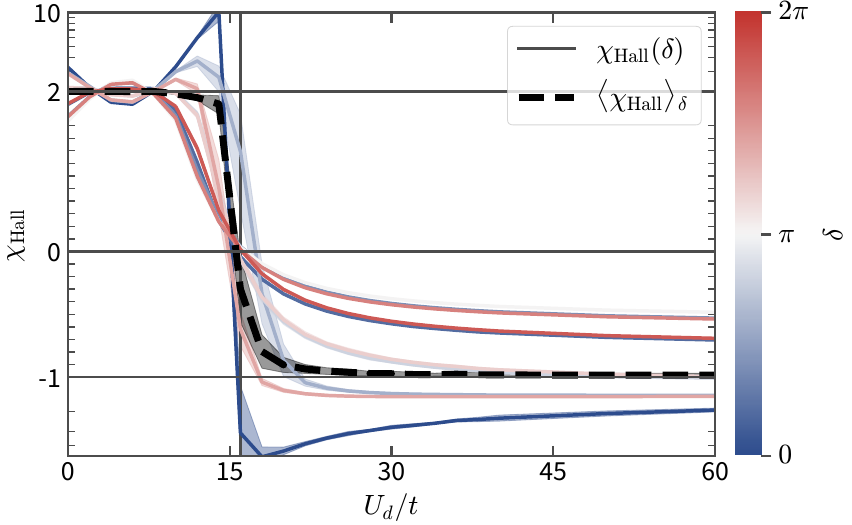}
	\mycaption{Hall response for an adiabatically applied potential $\V$ for $W=2$.}{
    Data are shown without off-diagonal interaction terms, i.e., $U_o=0$.
    Therefore, we expect to observe the topological transition known from 1d systems at $U_d\approx W\cdot 8t$.
    The thin colored lines represent data for different twist angles $\delta\in\{0,0.1,0.2,0.3,0.4,0.5,0.6,0.7,0.8,0.9,1\}\times\pi$.
    The dashed, black line is the average over these $\delta$ values.
    Close to the 1d quantum phase transition for $\delta=0$ at $U_d\approx 16t$,
    the conductivity diverges, such that an extrapolation in $L$ is necessary, 
    see \cref{fig:jy_finite_size}.
    The $\chi_{\rm Hall}$ axis is logarithmic for $|\chi_{\rm Hall}|>1$ in order to emphasize the values between $\chi_{\rm Hall}=C\in\{-1,0,2\}$
    and the behavior close to the transition.
	}
	\label{fig:jy_uo0_w2}
\end{figure}

In \cref{fig:jy_uo0_w2}, the Hall conductivity for a cylinder of width $W=2$ is shown for various interaction strengths $U_d$ and twist angles $\delta$.
We find that the Hall conductivity depends both on the twist angle $\delta$ and the interaction strength $U_d$.
The average $\langle\chi_{\rm Hall}\rangle_\delta$ shown in gray assumes the quantized values $C=2$ ($C=-1$) for weak (strong) interactions.
We cannot resolve the topological transition accurately due to the short length of the simulated systems.

For most values of $\delta$, 
the Hall response crosses $\chi_{\rm Hall}(U_d)=0$ continuously at the topological transition.
Even though the susceptibility is not quantized in a single, 
finite-size system,
we can observe the change of sign and amplitude of $\chi_{\rm Hall}$ associated with the topological transition from a single twist angle $\delta$.
The exception are values close to $\delta=0$, 
for which $\chi_{\rm Hall}$ diverges.
We discuss this in the next section.

\subsubsection{Divergence at the phase transition}
\label{sec:finite_l}
Without twist angle, $\delta=0$, 
and for $U_o=0$,
the $k=0$ leg corresponds to the  AB$_2$ ionic Hubbard model \cite{Yamamoto2001,Torio2006,MurciaCorrea2016}.
This 1d model exhibits two phase transitions as a function of the interaction strength: 
from a band insulator (BI), to a spontaneously dimerized insulator (SDI), to a correlated Mott insulator (MI) \cite{Fabrizio1999}.
For the parameters chosen in this paper, 
we cannot resolve both transitions 
because the critical values of the interaction strength $U_d$ are very close to each other
and much longer systems would be required \cite{Stenzel2019}.

In the intermediate SDI phase,
different dimer orientations create a two-fold ground-state degeneracy \cite{Fabrizio1999}.
For the ionic Hubbard model, this causes a diverging electric susceptibility \cite{Manmana2004,Tincani2009}, 
due to the different center-of-mass (COM) positions of both dimer configurations.
In the hybrid-space representation, 
the different COM positions along the $x$ direction for fixed quasimomentum $k=0$
correspond to different currents $\langle \hat j_y\rangle$,
see \cref{eq:hsjy}.

\begin{figure}[tb] 
	\centering \includegraphics[width=\columnwidth, clip]{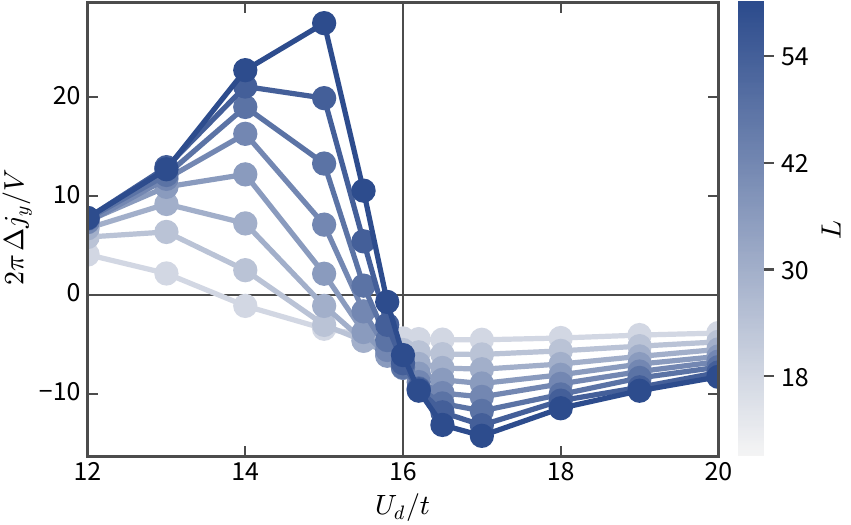}
	\mycaption{Finite-length dependence of the Hall current close to the topological phase transition.}{
	Data are shown for $W=2$ at $U_o=0$ and $\delta=0$ for $L\in\{18,24,30,36,42,48,54,60\}$.
	Unlike for other plots, 
	we do not perform a linear fit of $\chi_{\rm Hall}$
 	because the response can be nonlinear close to the transition.
	Here, we compute the Hall current as
	$\Delta j_y := j_y(V) - j_y(0)$, where $j_y=\langle \hat j_y\rangle$
	is computed for ground states and $V=2.8\cdot10^{-3}t$.
	For $L\to\infty$, 
    the Hall current diverges and changes its sign at the phase transition $U_d\approx 16t$, where the 1d superlattice model exhibits a spontaneously dimerized phase.
	For other twist angles $\delta$, 
	the response $\chi_{\rm Hall}$ does not diverge,
	but crosses through zero continuously, as shown in \cref{fig:jy_uo0_w2}.
	}
	\label{fig:jy_finite_size}
\end{figure}

In \cref{fig:jy_finite_size}, Hall currents close to the topological transition are shown for different system lengths.
To reduce the numerical cost, results are computed for $W=2$.
However, since the divergence of $\langle \hat j_y\rangle$ is only due to the $k=0$ leg,
increasing $W$ should not make a qualitative difference when the legs are uncoupled at $U_o=0$.

We find that for longer cylinders, 
the interaction strength $U_d$ at which $\chi_{\rm Hall}=0$ approaches the critical value $W\cdot U_{c,\mathrm{1d}}\approx 16t$ from below.
The diverging Hall response indicates a discontinuity in $\langle \hat j_y\rangle$ for $L\to\infty$.

\subsection{Thin cylinder limit}
\label{sec:w2}
In order to study a broad range of interaction strengths $U_d,\,U_o$,
we choose a width of $W=2$, 
which is the easiest to study numerically.
In the real-space representation, 
the case of $W=2$ seems to be special:
If there are only two legs, 
a particle cannot move around the ``cylinder'',
thus all complex tunneling rates vanish
and there is no flux, cf.\ \cite{Grusdt2014},
\begin{equation}
\label{eq:fhhw2}
\begin{aligned}
\hamil_{W=2}=\sum_{x,y}\bigg[
\bigg(&-t_y\cos(2\pi \alpha x-\delta/2)\VCh_{x,y}\cdot\VC_{x, y+1} \\
&-t\VCh_{x,y}\cdot\VC_{x+1,y} +\hc\bigg)        
+\U\bigg]\,.
\end{aligned}
\end{equation}
However, we argue that this is rather due to the chosen basis:
in the hybrid-space representation in \cref{eq:hshh}, 
there are no complex phases or tunneling along the $y$ direction, anyway.
We discuss the effect of a larger width in the following \cref{sec:width_scaling}.

\begin{figure}[tb] 
	\centering \includegraphics[width=\columnwidth, clip]{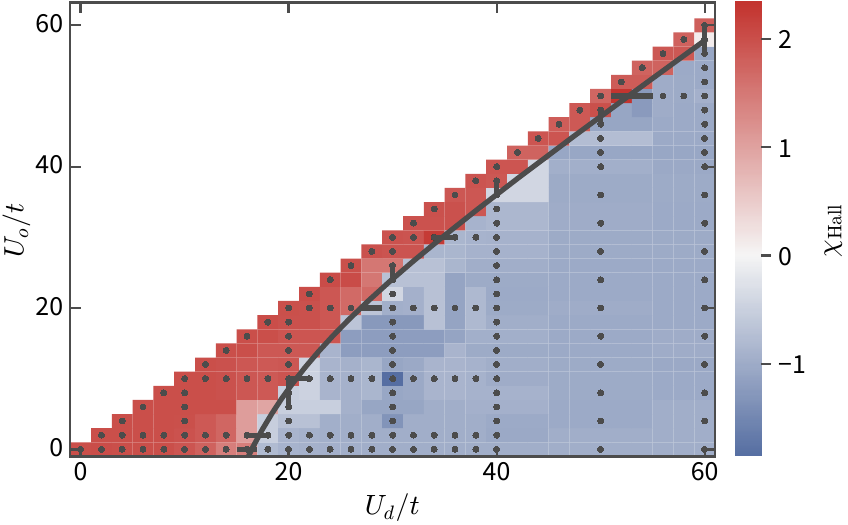}
	\mycaption{Topological phase diagram as a function of interaction strengths for $W=2$.}{
	The horizontal cut at $U_o=0$, corresponding to the 1d model, 
	is shown in \cref{fig:jy_uo0_w2}.
	We find that the $C=-1$ phase extends to finite values of the ring-wise interactions $U_o$.
	For large $U_d$,
	the topological transition approaches the diagonal $U_d=U_o$.
	However, the Hofstadter-Hubbard model ($U_d=U_o$) remains in the $C=2$ phase, 
	which is adiabatically connected to the band insulator,
	for all interaction strengths considered.
	A cut of this plot at $U_d=40t$ is shown in \cref{fig:jy_ud40_w2}.
	The gray line indicates the topological phase transition, 
	it is estimated from the shown data set.
	Data are shown for $L=30,\,W=2$ and averaged over $\delta\in\{0,0.2,\,0.4,\,0.6,\,0.8,\,1,\,1.2,\,1.4,\,1.6,\,1.8\}\times\pi$.
    We have computed $\chi_{\rm Hall}$ for the parameters indicated by gray dots, 
    in between those points, we use an interpolation for visualization purposes.
	}
	\label{fig:jy_ud_uo_phase_diagram}
\end{figure}

\Cref{fig:jy_ud_uo_phase_diagram} shows the Hall conductivity for various interaction strengths $U_d$ and $U_o$.
We find that the $C=-1$ phase extends to the region $U_o>0$ for strong interactions $U_d>16t$, 
depicted by the blue region.
As we further increase $U_d$, 
the $C=-1$ region becomes larger, such that it approaches the Hubbard-Hofstadter limit on the diagonal, at $U_d=U_o$.

The data in \cref{fig:jy_ud_uo_phase_diagram} are averaged over 10 values of the twist angle $\delta$.
This is not necessarily sufficient to verify integer quantization,
as one can see by the slight variations in color.
However, as the quantization is topological,
it suffices to verify integer values for single combinations of interaction strengths $U_d$ and $U_o$, 
as shown in \cref{fig:jy_quantization}.

The gray line in \cref{fig:jy_ud_uo_phase_diagram} shows our estimate of the phase boundary between $C=-1$ and $C=2$ phases. 
Up to $U=U_d=U_o=60t$,
the Hofstadter-Hubbard model seems to remain in the $C=2$ phase, 
which is adiabatically connected to the free model.
This result indicates that there is a topological phase transition between interacting 1d charge pumps 
and the interacting 2d Hofstadter Hubbard model.

\subsection{Transition in wider cylinders}
\label{sec:width_scaling}
As stated in \cref{sec:w2}, 
the case of $W=2$ seems to be different from wider cylinders.
While we expect the $C=-1$ phase to exist in the quasi-1d limit ($U_o=0$) for any system size,
the required interaction strength $U_d$ is proportional to the width $W$ due to the prefactor in \cref{eq:ud_uo}.
Thus, the $C=-1$ phase might not exist in the 2d thermodynamic limit.

\begin{figure}[tb] 
	\centering \includegraphics[width=\columnwidth, clip]{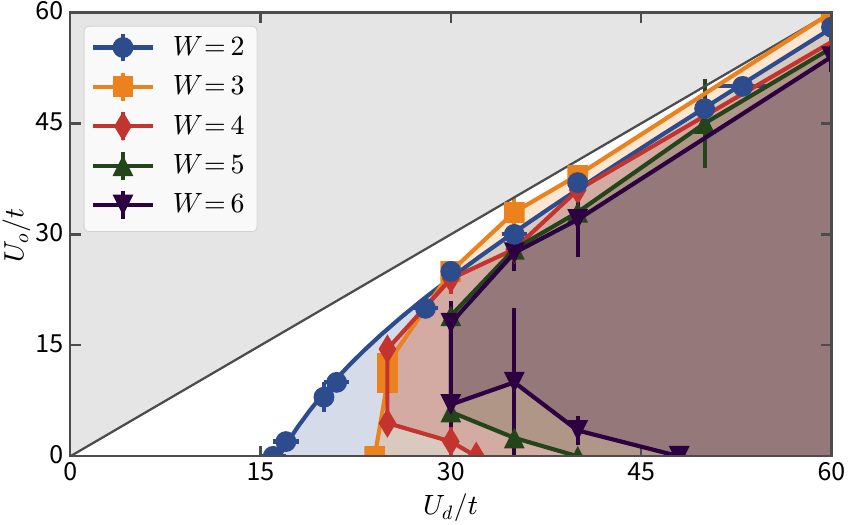}
	\mycaption{Topological transition for different widths.}{
		The colored regions indicate the $C=-1$ phase, 
		the lines are guides to the eye.
		Parameters in the gray region, $U_o>U_d$ have not been considered for this plot.
		For $U_o=0$ the critical interaction strength $U_{d,c}$ scales proportional to $W$,	
		for finite $U_o$ the dependence on width decreases. 
        Data were obtained for length $L=30$. 
        For $W=2$, we use data from \cref{fig:jy_ud_uo_phase_diagram}, 
        averaging 10 twist angles $\delta$, 
        results for wider cylinders were computed only for $\delta=\pi$.
	}
	\label{fig:topo_transition_width}
\end{figure}

In \cref{fig:topo_transition_width}, we show the boundary of the $C=-1$ phase for $U_o<U_d$ at widths $W=2,3,4,5,6$.
The data are obtained from a single value of $\delta=\pi$ such that we can measure the sign of the response,
but $\chi_{\rm Hall}(\delta)$ is not quantized, 
cf.\ \cref{fig:jy_uo0_w2}.

We observe that the shape of the phase boundary changes with width:
For $W\geq 4$, there exist regions of $C=-1$ at smaller $U_d$ than what we would expect from scaling up $W=2$ data, i.e., 
$U_{d, c}(U_o>0)< W\,U_{c, \mathrm{1d}}$.

The data in \cref{fig:topo_transition_width} might indicate that parts of the phase boundary do not change with $W$.
Close to $U_d=25t,\; U_o=15t$, there might be a point where the phase boundaries for $W=2,3,4$ coincide.
However, we could not obtain reliable data for $W=5,6$ to confirm this observation.
If any part of the phase boundary is independent of the width, 
the $C=-1$ phase will also exist in the 2d thermodynamic limit for finite $U_d$ and $U_o$.

\section{Ferromagnetic ground state}
\label{sec:fm_gs}
In the previous \cref{sec:jy_s0}, we have restricted the DMRG algorithm to the $S=0$ spin-singlet symmetry sector.
The singlet is the lowest energy state, both for the 1d superlattices \cite{Stenzel2019} and for the 2d Hofstadter-Hubbard model.
However, 
%\st{for a region between both models,
%	the ferromagnetic $S=N/2$ sector is the lowest in energy.
%}
%\hl{
	for some parameters in the $C=-1$ phase, 
	we find spin sectors with $S>0$ to be the lowest in energy.
	In particular, the true ground state can be in the ferromagnetic (FM) sector with $S=N/2$.
	The dependence of energy on total spin $S$ is further discussed in \cref{sec:S_sectors}.
%}

\subsection{Width \texorpdfstring{$W=2$}{W=2}}

\begin{figure}[tb] 
	\centering \includegraphics[width=\columnwidth, clip]{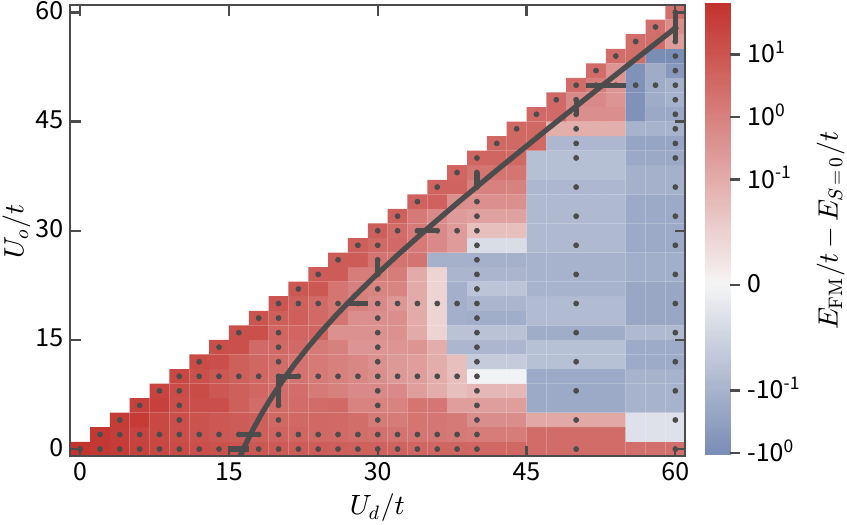}
	\mycaption{Energy difference between the lowest energy ferromagnetic state and the ground state in the spin-singlet sector.}{
	The spin singlet is the true ground state both for the 1d superlattice model 
	and the 2d Hofstadter model.
	Data correspond to the systems shown in \cref{fig:jy_ud_uo_phase_diagram}, 
    the gray line indicates the topological transition that we show in that plot.
    The gap is averaged over twist angles $\delta$ and computed for $L=30$ and $\;W=2$ for the interaction parameters indicated by the gray dots.
    The shading is interpolated for visualization purposes.
	}
	\label{fig:fm_gap_ud_uo_phase_diagram}
\end{figure}

In \cref{fig:fm_gap_ud_uo_phase_diagram}, we show the energy difference between the ground-state energy in the ferromagnetic sector $E_{\rm FM}$ and the lowest energy spin-singlet state $E_{S=0}$.
Depicted by the blue region, 
there exists a FM region for strong interactions $U_d\gtrsim 40t$ and finite, 
but smaller interaction strength $0<U_o<U_d$.
Deep in the red, spin-singlet (blue, FM) region, 
the energy increases (decreases) monotonically as a function of total spin $S$.
At the boundary, energy sectors with $0<S<N/2$ can be energetically favorable.
The precise position of $E_{\rm FM}=E_{S=0}$ also depends on the twist angle $\delta$.

In \cref{fig:fm_gap_ud_uo_phase_diagram}, we also show the gray line depicting the topological phase boundary from \cref{fig:jy_ud_uo_phase_diagram}.
The region with the FM ground state lies entirely inside the $C=-1$  phase.

For a FM state, we would indeed expect a Chern number $C=-1$:
Double occupation is prohibited by Pauli's principle, 
both in real space and hybrid space.
Therefore, both $\langle\Ud\rangle$ and $\langle\U\rangle=\langle \Uo+\Ud\rangle$ vanish, 
and the spatial component of the wave function equals that of free, 
spinless fermions.
A single species of fermions at particle density $\rho=2/3$ would occupy the lowest two bands of the Hofstadter model, 
such that the total Chern number would be the sum of the lowest two bands, $C=1-2=-1$.

%\hl{
The numerical results shown in \cref{sec:S_sectors} do not exhibit any dependence of the Hall response $\chi_{\rm Hall}$ on total spin $S$, when the ground state is in the FM region.
%}

\subsection{Existence for wider cylinders}
\begin{figure}[tb] 
	\centering \includegraphics[width=\columnwidth, clip]{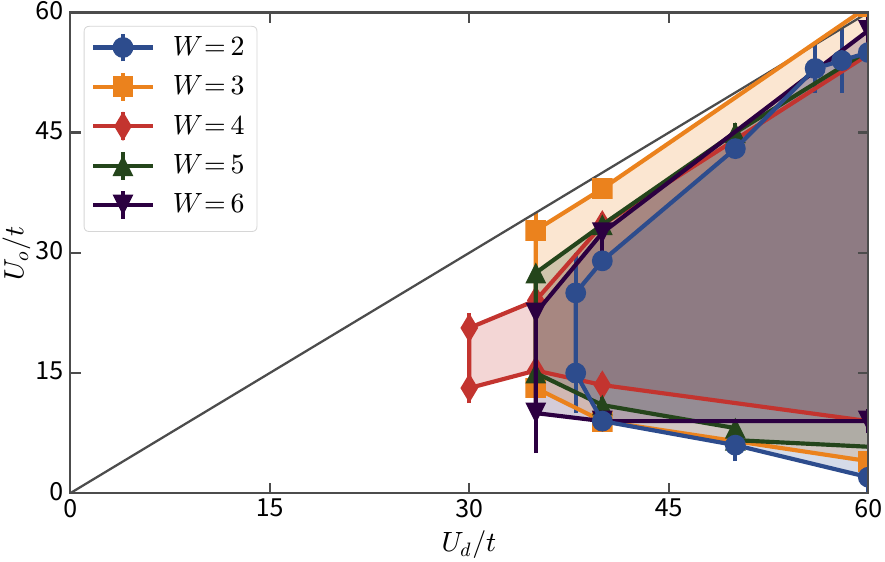}
	\mycaption{Region with a ferromagnetic ground state for different widths.}{
	The shaded region indicates where the ferromagnetic state is lower in energy than the spin singlet, 
	i.e., the blue area in \cref{fig:fm_gap_ud_uo_phase_diagram}.
	The lines are only guides to the eye, 
	error bars indicate the step size used for the interaction strength $U_o$.
	Data are shown for $L=30,\,\delta=\pi$, except for $W=2$, 
	which is averaged over 10 values of $\delta$ as in \cref{fig:fm_gap_ud_uo_phase_diagram}.
	The boundaries do not seem to change strongly when going to wider systems.
	We expect the dependence on $W$ to be smaller, 
	when an average over the twist angle $\delta$ is also taken into account.
	}
	\label{fig:FM_transition_width}
\end{figure}

In \cref{fig:FM_transition_width}, we show how the extent of the FM ground state changes for wider cylinders.
The boundary does not seem to change significantly as the system gets wider.
Some fluctuations have to be expected,
because the boundary also depends on the twist angle $\delta$ 
and going to larger $W$ effectively changes $\delta$.

This result seems to indicate that the FM phase also exists for large systems at finite $U_d$ and $U_o$.
If the appearance of the FM phase is related to the fact that we observe $C=-1$ in the spin-singlet state,
this would suggest that the $C=-1$ phase also exists for larger systems at finite $U_d$ when $U_o>0$.

We note that DMRG tends to overestimate the extent of the ferromagnetic ground state in the $U_o,\, U_d$ diagram,
especially for wide cylinders:
The energies $E_{S=0}$ are an upper bound to the true value,
while we compute $E_{\rm FM}$ numerically exactly.

\section{Summary}
\label{sec:summary}
We studied the fermionic Hofstadter model numerically on a cylinder,
in a hybrid-space representation.
We considered tuneable interactions such that on-site repulsion in hybrid space (1d superlattice limit)
and on-site repulsion in real space (2d Hubbard-Hofstadter limit) are the limiting cases.
This parameterization allows us to connect interacting 1d charge pumps to interacting 2d Chern insulators. 

For weak interactions, 
the 1d and 2d models are adiabatically connected to the same free model,
thus, they exhibit the same topological properties.
The 1d model is known to undergo quantum phase transitions for strong interactions \cite{Fabrizio1999,Manmana2004,Torio2006,Torio2001,Otsuka2005,Kampf2003,Batista2004,Lou2003,Aligia2004,MurciaCorrea2016,Yamamoto2001,Egami1993},
changing its topological properties \cite{Stenzel2019,Hu2019}.

In the quasi-1d case, where the hybrid-space legs are uncoupled, 
we reproduced the interaction-driven topological transition from a $C=2$ topological insulator to one with Chern number $C=-1$.
Depending on system size, averaging the Hall currents over twisted boundaries may be necessary to show topological quantization.

The interacting $C=-1$ insulator is robust under changes of the interaction strength.
In our parameterization, it almost reaches the 2d Hubbard-Hofstadter limit.
We verified the existence of the interacting $C=-1$ phase for numerically accessible cylinder widths $W\in\{2,3,4,5,6\}$, 
and found that it extends to larger parameter regions than we would expect from scaling up data for $W=2$.

We computed the Hall response directly
by applying a weak potential gradient adiabatically.
Similar setups have already been realized in experiments with ultracold atoms \cite{Aidelsburger2015,Stuhl2015,Mancini2015}.
We showed that we can measure an integer-quantized Hall response even for strongly-interacting systems.
Our approach relies on periodic boundaries along the width of the system, 
which may be realizable in synthetic dimensional lattices \cite{Celi2014}.
However, weak quenches in open systems should yield similar results.

We also observed a region between 1d and 2d Hubbard interaction, 
where a ferromagnetic (FM) state is lower in energy than the spin-singlet sector.
This region lies entirely inside the $C=-1$ phase.
We showed that the FM phase exists for all widths considered.
The phase boundary does not seem to depend strongly on the width $W$,
indicating that the FM phase is robust for larger systems.
A FM ground state necessarily has Chern number $C=-1$,
due to the bandstructure of the Hofstadter model.
This may indicate that the $C=-1$ phase in the spin-singlet symmetry sector is related to the FM ground state.
Putting this observation onto firmer grounds is left for future research.

%\hl{
All numerical results were obtained for a model with anisotropic tunneling rates, $t_y=1.5t$.
Additional data (not shown here) for the isotropic case of $t_y=t$ show qualitatively similar results:
Both the $C=-1$ phase and ferromagnetism exist in the $W=2$ limit.
The role of anistropic tunneling rates remains an interesting question, 
see also other recent studies of the Hofstadter model \cite{Koshino2004,Hugel2017,Mastropietro2019}.
%}

The family of models studied in this paper is clearly motivated from theoretical considerations.
However, tuneable on-site and leg-wise interactions can be realized in synthetic-dimensional lattices \cite{Tanzi2018}.
While our results show that two-leg ladders suffice to observe a topological transitions,
further research on more readily realizable models is necessary.

We thank M. Buser and C. Hubig for useful discussions.
This research was funded by the Deutsche Forschungsgemeinschaft
(DFG, German Research Foundation) via Research Unit FOR 2414
under project number 277974659.
U.S. acknowledges support by the Deutsche Forschungsgemeinschaft (DFG, German Research Foundation) under Germany's Excellence Strategy EXC-2111-390814868.
\appendix

\section{Additional plots of Hall response}
\label{sec:app_chi}
\begin{figure}[tb] 
	\centering \includegraphics[width=\columnwidth, clip]{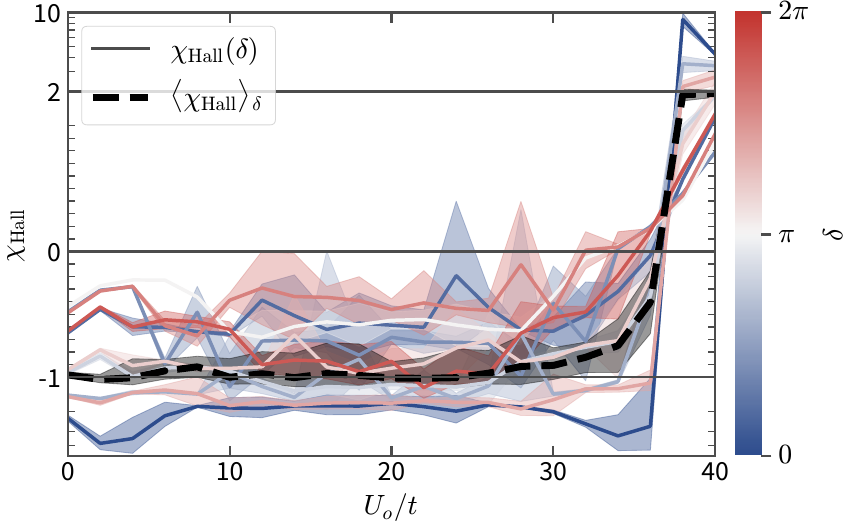}
	\mycaption{Hall response for a cut through \cref{fig:jy_ud_uo_phase_diagram} at $U_d=40t$.}{
		Thin, colored lines show the fitted value of $\chi_{\rm Hall}$,
		the corresponding shaded regions indicate the uncertainty as measured by the cost of the fit:
		Higher cost corresponds to less linear behavior of $\langle \hat j_y\rangle (V)$, 
		due to numerical errors and finite-size effects.
		Errors are larger than in \cref{fig:jy_uo0_w2},
		because the terms of $\Uo$ greatly increase the numerical complexity.
		The average over all $\delta$ values is shown as dashed, black line.
		It takes the value $\chi_{\rm Hall}=-1$ up to $U_o=30t$ and $\chi_{\rm Hall}=2$ for $U_o\gtrsim 37t$. 
		Data are show for $W=2,\,L=30,\,\delta\in\{0,0.2,0.4,0.6,0.8, 1,1.2,1.4,1.6,1.8\}\times\pi$.
		Greater numerical precision and more samples would be required to show integer quantization.
		The $\chi_{\rm Hall}$ axis is logarithmic for $|\chi_{\rm Hall}|>1$,
		in order to suppress the outliers for $\delta=0$, see \cref{sec:finite_l}, 
		and to focus on the topological transition.
	}
	\label{fig:jy_ud40_w2}
\end{figure}

In this section, we show plots for the Hall conductivity $\chi_{\rm Hall}$,
complementary to \cref{fig:jy_uo0_w2} and \cref{fig:jy_ud_uo_phase_diagram} in the main text.
In \cref{fig:jy_ud40_w2}, 
we show a cut through \cref{fig:jy_ud_uo_phase_diagram} for finite $U_o\geq 0$
and fixed $U_d=40t$.
We observe the topological phase transition from $C=-1$ to $C=2$ close to $U_o=36t$.
The sign of the response $\chi_{\rm Hall}$ does not depend on the twist angle $\delta$,
except close to the transition, 
where a finite-size extrapolation would be required, cf.\ \cref{sec:finite_l}. 
The errors of the fits are larger than in \cref{fig:jy_uo0_w2} 
because the increased number of terms in $\Uo$ makes the problem numerically harder.

\begin{figure}[tb] 
	\centering \includegraphics[width=\columnwidth, clip]{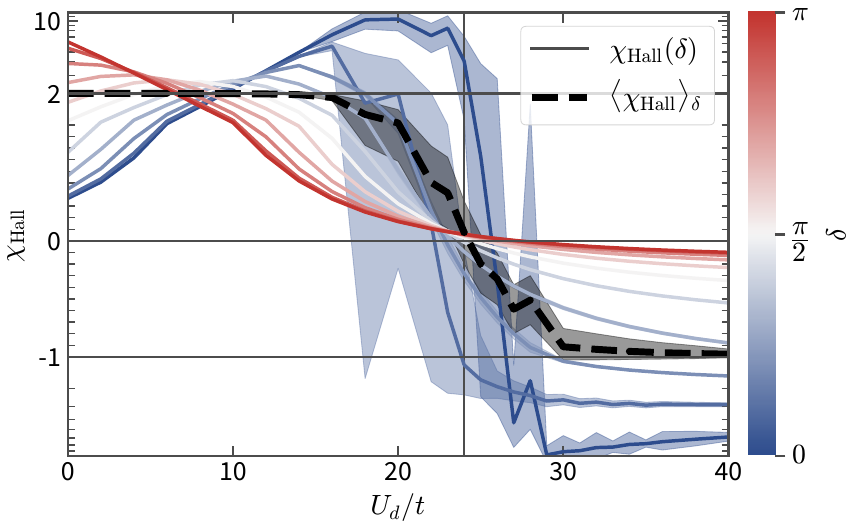}
	\mycaption{Hall conductivity across the quasi-1d phase transition, when $U_o=0$, for $W=3,\,L=24$.}{
		The thin colored lines are data for different twist angles $\delta\in\{0,0.1,0.2,0.3,0.4,0.5,0.6,0.7,0.8,0.9,1\}\times\pi$.
		The dashed, black line is the average over these values of $\delta$.
		Close to the 1d quantum phase transition for $\delta=0$ at $U_d\approx 8t\,W$,
		the conductivity diverges such that an extrapolation in $L$ is necessary, 
		see \cref{fig:jy_finite_size}.
		The $\chi_{\rm Hall}$ axis is logarithmic for $|\chi_{\rm Hall}|>1$, 
		in order to suppress outliers for $\delta=0$, see \cref{sec:finite_l}, 
		and to focus on the topological transition.
	}
	\label{fig:jy_uo0_w3}
\end{figure}

We show a plot for the Hall response in the quasi-1d case, 
when $U_o=0$, in \cref{fig:jy_uo0_w3}.
This plot corresponds to \cref{fig:jy_uo0_w2}, 
but for width $W=3$, 
which is commensurate with the magnetic unit cell at $\alpha=1/3$.
As expected, we observe the transition from $C=2$ to $C=-1$ at $U_d\approx 8Wt$.
We cannot resolve the behavior of $\chi_{\rm Hall}$ at the phase transition for $\delta\approx 0$.
For other values of $\delta$, 
the error of $\chi_{\rm Hall}$ is small
and the curves are smooth,
even at the point where $\chi_{\rm Hall}$ changes sign.
In the quasi-1d case, the Hall response for systems of different widths can be related via
\begin{multline}
	\chi_{\rm Hall}(2W,\delta,2U_d,U_o=0) =\big[\chi_{\rm Hall}(W,\delta,U_d,U_o=0)\\ +\chi_{\rm Hall}(W,\delta+\pi,U_d,U_o=0) \big]/2\,.
\end{multline}
We verified this relation numerically with simulations for width $W=6$.

\section{Energy based filtering}
\label{sec:e_v_fit}
As described in \cref{sec:method}, 
estimating the error of DMRG results is generally difficult, 
especially for observables other than the energy.
We are primarily concerned with errors of the Hall current $\langle \hat{j}_y\rangle$:
We compute the current for $5$ to $10$ values of the linear potential $0\leq V\leq 10^{-2}t$.
The difference of measured currents $\langle \hat{j}_y(V+\Delta_V)\rangle - \langle \hat{j}_y(V)\rangle$ is thus on the order of $10^{-3}t$,
such that slight convergence issues can drastically affect the quality of the results.
This section describes our method to control the  convergence of DMRG simulations.

When studying numerically challenging system sizes,
we use a method to filter DMRG results,
which were obtained for different strengths of the linear potential $V$.
Since $V\ll t$ is small and $\V$ is positive semidefinite, 
we assume a linear response in energy, $E_{\rm gs}(V) = E_{\rm gs}(0) + c\,V$, for some non-negative number $c$.
DMRG is a variational method
and therefore, we can estimate the true ground-state energy $E_{\rm gs}(V)$ by fitting a \emph{lower}, linear envelope to the numerical data.

\begin{figure}[tb] 
    \centering \includegraphics[width=\columnwidth, clip]{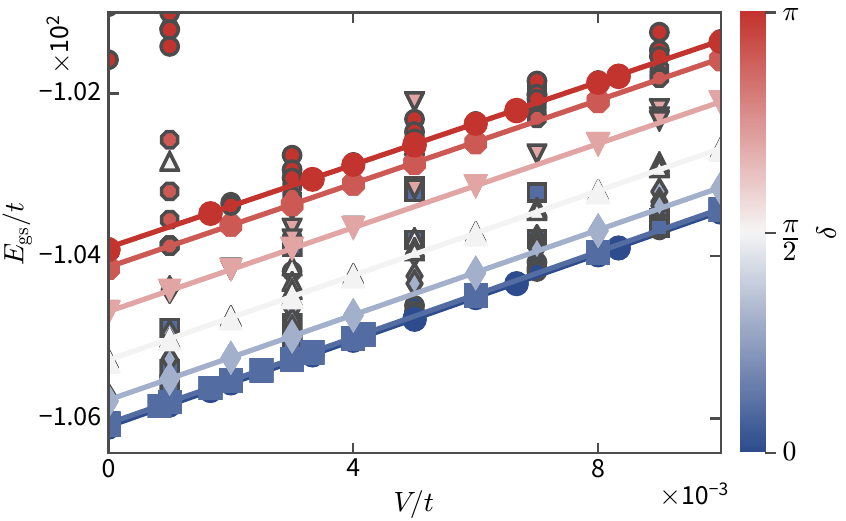}
    \mycaption{Filtering DMRG results by comparing variational energies.}{
    The data are shown for $L=12,\,W=6,\,U_d=18t,\,U_o=0$.
    Markers represent variational DMRG results for different values of twist angle $\delta$.
    The lines are a linear fit to the lower envelope of the DMRG data.
    Markers with gray outline lie above the fit by some threshold $\delta_E$,
    and are thus discarded.
    In this plot, 
    we choose $\delta_E\approx 10^{-4}t$, 
    such that only extreme outliers can be seen with the bare eye.
	}
    \label{fig:energy_fit}
\end{figure}

In \cref{fig:energy_fit}, we show such fits for different twist angles $\delta$.
We then ignore data from DMRG states,
which have energies above the fit, by some threshold.
The plot only illustrates the method rather than showing its result 
because the threshold is too small to see all discarded states.

The data for \cref{fig:energy_fit} are obtained by reusing previous MPS:
To compute a state for $V+\Delta_V$, 
we use the truncated MPS for a potential of strength $V$ as the initial state.
However,
multiple runs with different random states for $V=0$ and different step sizes $\Delta_V$ have been used.

\section{Quench dynamics}
\label{sec:quench_dynamics}
In \cref{sec:jy_s0}, we compute the Hall response $\chi_{\rm Hall}$ adiabatically,
meaning that we perform DMRG sweeps for each value of the potential strength $V$.
Numerically, this is a rather cheap approach, 
requiring data for only a few values of $V$ to obtain quantitative results.

\begin{figure}[tb] 
    \centering \includegraphics[width=\columnwidth, clip]{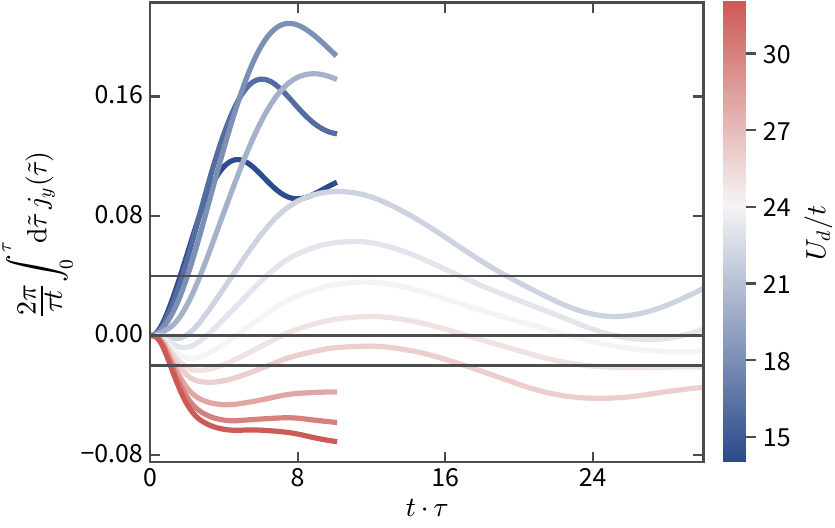}
    \mycaption{Time-dependent response of the Hall current $\langle \hat j_y\rangle$ after quenching a linear potential from $V=0$ to $V=0.02t$.}{
    The data are shown for $L=18,\,W=3,\,\delta=0,\,U_o=0$ for different interaction strengths $U_d$.
    We average the current over time $\tau$ to suppress oscillations.
    The gray, horizontal lines indicate values for $C\in\{-1,0,2\}$.
    For a finite system size, we would have to average over several twist angles $\delta$
    in order to observe quantization 
    and the topological transition at $U_d\approx 24t$.
    However, there is a change of the sign and amplitude of $\langle \hat j_y\rangle$ as we cross this transition.
    We restrict the simulation time to $\tau\,t\leq 10$ for small and large $U_d$,
    because there was no ambiguity in the sign of $\langle \hat j_y\rangle$.
    }
    \label{fig:quench_dynamics}
\end{figure}

In an experiment, it might be easier to prepare the ground state for $V=0$
and to observe its evolution upon quenching a weak potential $0\neq V \ll t$.
In \cref{fig:quench_dynamics}, 
we show that the change of the Chern number can also be measured in such quench experiments.
While the system size shown in \cref{fig:quench_dynamics} is too small to observe quantization, both the 
sign and the amplitude of $\chi_{\rm Hall}$ change as the interaction strength crosses the critical value $U_d\approx 24t$.
We show data for a single twist angle $\delta$,
since averaging over twist angles might not be possible in experiments, either.

In order to probe the regime of linear response, 
we switch on a weak potential $t\gg V>0$, 
such that the state remains ``close'' to the ground state.
Therefore, the entanglement entropy does not increase strongly,
and rather long times $\tau\,t > 10$ can be reached at small bond dimensions.

The data in \cref{fig:quench_dynamics} is obtained using a single-site variant of the TDVP algorithm \cite{Haegeman2016,Paeckel2019}.
We use a step size of $\Delta_\tau t = 0.1$ 
and fix the bond dimensions at $m_{SU(2)} =3000$.
We verify the results up to $\tau t=10$ by comparing with other simulations:
There is good agreement with the two-site TDVP method 
and with the result of simulations performed with $\Delta_\tau t=0.05$ as well as $m_{SU(2)}=5000$.

\section{Hall response in different spin \texorpdfstring{$S$}{S} symmetry sectors}
\label{sec:S_sectors}
%\hl{
All DMRG simulations in the main text have been performed in the spin-singlet symmetry sector.
The comparison with the FM ground state in \mbox{\cref{sec:fm_gs}} does not require DMRG
because both interaction terms $\Uo$ and $\Ud$ vanish for any FM state.
Therefore, the FM ground state always has Chern number $C=\langle \chi_{\rm Hall}\rangle_\delta = -1$, 
regardless of the interaction strength.

\begin{figure}[tb] 
	\centering \includegraphics[width=\columnwidth, clip]{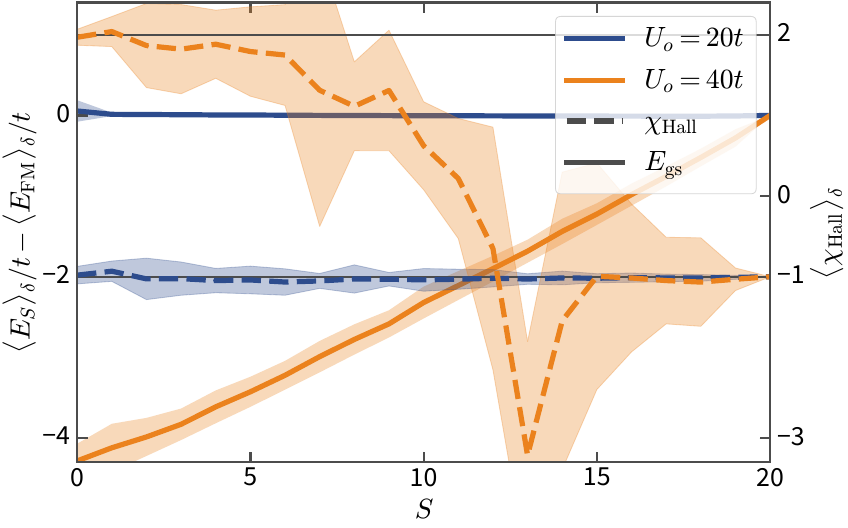}
	\mycaption{Ground-state energy and Hall response in different symmetry sectors of the total spin $S$.}{
	Data are displayed for $L=30,\, W=2$ and $U_d=40t$ and are averaged over $\delta\in \{0,0.2,0.4,0.6,0.8,1,1.2,1.4,1.6,1.8\}\times\pi$.
	We show data for different values for $U_0$:
	For $U_o=20t$, we are in the $C=-1$ phase.
	In this case, the Hall response agrees with $C=-1$ for all values of $S$
	and the FM state ($S=20$) is slightly lower in energy than the spin singlet.
	For $U_o=40t$, we find $\langle\chi_{\rm Hall}\rangle_\delta = C=2$ in the spin-singlet ground state.
	Increasing $S$ leads to an increase of the ground-state energy but the Hall response remains consistent with $C=2$ up to $S=10$.
	For larger $S$, we recover the $C=-1$ phase since all interaction terms vanish for a FM state.
	For the ground-state energy, 
	the shaded region is the standard deviation with respect to $\delta$.
	}
	\label{fig:e_chi_over_spin}
\end{figure}

To elucidate the dependence of the ground-state energy and Hall response on total spin $S$,
we show numerical data for two interaction strengths in \mbox{\cref{fig:e_chi_over_spin}}.
For the parameters $U_d=40t$ and $U_o=20t$, 
the spin singlet yields $C=-1$ and 
$E_{\rm FM}<E_{S=0} $ such that we are in the FM phase as discussed in \mbox{\cref{sec:fm_gs}}.
Computing ground states for all other possible spin multiplets,
we find that the state for $S=18$ is actually the true ground state for this interaction strength.
However, the states are nearly degenerate with $E_{\rm FM}-E_{S=18}$ being on the order of $10^{-4}t$.
For stronger interaction $U_d$, 
the FM state is the true ground state, but the sectors remain nearly degenerate.
Our results for the Hall response $\langle \chi_{\rm Hall}\rangle_\delta$,
which we average over ten values for $\delta$,
do not depend on spin.
They agree with the Chern number $C=-1$ for all values of $S$.

For $U_d=U_o=40t$, the energy increases monotonically in $S$
and the spin singlet is the true ground state.
Since the singlet state is in the $C=2$ phase and the FM state has $C=-1$,
the Chern number can, in general, not be independent of $S$.
Our data suggest that the Hall response $\langle \chi_{\rm Hall}\rangle_\delta$ deviates from $C=-1$ and $C=2$ for $N/8<S<3N/8$.
The breakdown of quantization is plausible
because in the free model, the topological invariant is only well-defined when either $S=0$ or $S=N/2$.
%}

\bibliographystyle{apsrev4-1}
\bibliography{fermi-hofstadter-hubbard}

\end{document}